\documentclass[aps,prd,floatfix,superscriptaddress,showpacs,notitlepage,10pt]{revtex4-1}

\usepackage{graphicx}
\usepackage{dcolumn}
\usepackage{bm}
\usepackage{epsfig}

\newcommand\eqn[1]{Eq.\,(\ref{#1})}

\newcommand\fig[1]{Fig.\,{\ref{#1}}}
\newcommand\sect[1]{Sect.\,{\ref{#1}}}

\newcommand{\beq}{\begin{equation}}
\newcommand{\eeq}{\end{equation}}
\newcommand{\bea}{\begin{eqnarray}}
\newcommand{\eea}{\end{eqnarray}}

\newcommand{\nn}{\nonumber\\}

\begin{document}
\title{Regulator dependence of fixed points in quantum Einstein gravity with $R^2$ truncation}

\author{S. Nagy$^a$, B. Fazekas$^b$, Z. Peli$^a$, K. Sailer$^a$, I. Steib$^a$\\
$^a$Department of Theoretical Physics, University of Debrecen, P.O. Box 5, H-4010 Debrecen, Hungary\\
$^b$ Institute of Mathematics, University of Debrecen, P.O. Box 12, H-4010 Debrecen, Hungary}
\date{\today}

\begin{abstract}
We performed a functional renormalization group analysis for the quantum Einstein gravity including a quadratic term in the curvature. The ultraviolet non-gaussian fixed point and its critical exponent for the correlation length are identified for different forms of regulators in case of dimension 3. We searched for that optimized regulator where the physical quantities show the least regulator parameter dependence. It is shown that the Litim regulator satisfies this condition. The infrared fixed point has also been investigated, it is found that the exponent is insensitive to the third coupling introduced by the $R^2$ term.
\end{abstract}

\maketitle

\section{Introduction}

A very promising candidate of unifying the quantum theory and gravity is the so-called quantum Einstein gravity (QEG) \cite{Reuter:1996cp,Reuter:2007rv,Reuter:2012id}. The model is usually investigated in the framework of the functional renormalization group (RG) method \cite{Wetterich:1992yh,Morris:1993qb,
Berges:2000ew,Polonyi:2001se}. The RG technique enables us to compare the models at different energy scales $k$. It provides a systematic treatment of the quantum fluctuations starting from the high energy ultraviolet (UV) version of the model towards the low energy infrared (IR) one. The action is integrated out for all possible field configurations. In QEG the metrics plays the role of the field variable. In the simpest form of the model, i.e. by using the Einstein-Hilbert (EH) truncation of the action we have two couplings, the cosmological constant and the Newton constant. Besides their extreme importance in gravity their names are quite deceptive, since they are not constants, they depend on the RG scale $k$.

In perturbative sense QEG is non-renormalizable, because the Newton constant is irrelevant, therefore the model is not asymptotically free. This classification is valid only at the interaction free theory, which corresponds to the origin of the phase space spanned by the couplings, i.e. around the Gaussian fixed point (GFP). We usually throw out the irrelevant couplings in order to make our model renormalizable, but this is not possible in QEG, since the Newton constant plays a fundamental role in the description of gravitational interactions. One of the greatest result of the RG method in the past two decades was to find a UV non-gaussian fixed point (NGFP) in QEG. The genuine fixed point is UV attractive that makes the physical variables finite at arbitrarily high scale $k$. The UV NGFP makes the model asymptotically safe \cite{Percacci:2007sz,Reuter:2012xf,Nagy:2012ef,Gies:2013pma}. Much efforts are devoted to show that the UV NGFP exists if we include matter fields into the action, e.g. fermions \cite{Eichhorn:2011pc,Eichhorn:2011ec,Dona:2012am,Eichhorn:2016vvy}, scalar fields \cite{Percacci:2002ie,Percacci:2003jz,Zanusso:2009bs,Manrique:2010mq,Manrique:2010am,Vacca:2010mj,Eichhorn:2012va,Dona:2015tnf}, gauge fields \cite{Folkerts:2011jz,Eichhorn:2011gc,Donkin:2012ud,Christiansen:2017qca,Eichhorn:2017eht}, or even ghost sectors \cite{Groh:2010ta,Eichhorn:2010tb}.

Another plausible extension of QEG can be to go beyond the EH truncation of the action and to consider higher order terms in the scalar curvature $R$. There are works that treat the $f(R)$ truncation directly \cite{Codello:2007bd,Machado:2007ea,Dietz:2012ic,Gonzalez-Martin:2017gza}. It is already shown that the full flow equation for QEG including the quadratic term in the curvature $R$ also exhibits an UV NGFP \cite{Rechenberger:2012pm}. This pioneering work gave us the possibility for further investigations. Our goal is to show that the UV NGFP exists for further regulators in the $R^2$ truncation. We should use a regulator in the RG technique in order to remove divergences in the loop integrals. Besides, the regulator is an artificial term in the action, and every observable becomes regulator dependent after the unavoidable truncations and approximations in the RG calculations. Our goal is to perform a simple optimization program for the regulators, i.e. we look for that regulator, which exhibits the least sensitivity to the regulator parameters \cite{Litim:2001fd,Canet:2002gs}. We use the compactly supported smooth (css) regulator \cite{Nandori:2012tc,Nagy:2013hka}, which limits can provide us various regulators traditionally used in the RG area.

We calculated earlier the value of the critical exponent $\nu$ of the correlation length $\xi$ as the function of the css regulator parameters in QEG with EH truncation. We realized that a modified version of the Litim regulator shows the least sensitivity on the parameters \cite{Nagy:2013hka}. Here we repeat our investigation for QEG containing the $R^2$ term. First we map out the phase structure of the model with certain parameters of the css regulator. Then we look for the value of $\nu$ around the fixed points, and investigate how they depend on the regulator parameters. In \sect{sect:mod} we briefly present the model of QEG with the $R^2$ term and the regulators. The results for the UV and the IR regimes are summarized in \sect{sect:res} and the conclusions are drawn up in \sect{sect:con}.

\section{The model}\label{sect:mod}

The gravitational sector of the effective action for QEG up to $R^2$ terms can be written as
\bea 
\Gamma_k^{\text{grav}} = \int d^d x \sqrt{g} \biggl[\frac{1}{16 \pi G_k}\biggl(-R + 2\bar\lambda_k \biggr)+ \frac{1}{\bar b_k}R^2 \biggr],
\eea
where $G_k$ is the Newton constant, $\bar\lambda_k$ is the cosmological constant, and $\bar b_k$ is the coupling belonging to the $R^2$ term. The derivation of the evolution equations for the couplings is available in \cite{Lauscher:2002sq} for arbitrary dimension $d$ and regulator ${\cal R}_k$. In \cite{Rechenberger:2012pm} the Litim regulator was used, where the loop integrals in the $\beta$ functions can be performed analytically. In order to map out the phase structure of the model we introduce the dimensionless couplings according to
\beq
g = \frac{G_k}{k^{2-d}},~~\lambda = \frac{\bar\lambda_k}{k^2},~~b = \frac{\bar b_k}{k^{4-d}}.
\eeq
The $\beta$ functions define several singular locus due to the possible divergence of either the anomalous dimensions or the denominators which tend zero for certain trajectories. The latter singularity is usually related to the IR fixed point in the broken symmetric phase. It might happen that the locus separates in space the UV NGFP and the GFP for certain choices of ${\cal R}_k$. We consider $d=3$, because the phase structure is simpler than in $d=4$, e.g. in the former case we have a single UV NGFP, while we have two pieces of them in the latter dimension \cite{Rechenberger:2012pm}. 

The regulators play prominent role in the RG treatment, since they regularize the appearing UV or IR infinities in the flow equations. Roughly speaking we modify the dispersion relation by adding a momentum (and RG scale) dependent term into the action. It works as if we introduced new elementary excitations or quasi particles into the theory. The deviation from the original excitations makes unavoidable to investigate the regulator dependence, since different regulators can define different quasi particles, and we should know whether the observables are sensitive to such modifications, or not.

There are several commonly used regulators. Among them the Litim regulator is the most popular, because it is optimal in local potential approximation (LPA). Furthermore it provides analytic flow equations, since the loop integral in the Wetterich equation can be analytically calculated. However there is a need to investigate the regulator dependence. There are several possibilities to choose further regulators \cite{Pawlowski:2015mlf}. We chose the css regulator \cite{Nandori:2012tc,Nagy:2013hka} which has the form
\beq\label{eq:cssreg} 
r_{css} = \frac{s_1}{\exp[s_1 y^c/(1-s_2 y^c)]-1} \theta(1-s_2 y^c),
\eeq
where we choose $c=1$. We introduced $r={\cal R}_k/p^2$ and $y=p^2/k^2$, where $p$ is the momentum of the loop integrals in the Wetterich equations. Here we use a simplified css regulator form, which has only two free parameters $s_1$ and $s_2$ \cite{Nagy:2013hka}. Certain limits of the regulator in $s_1$ and $s_2$ can give us back some traditionally used regulators, i.e.
\bea\label{limregs}
\lim_{s_1\to 0} r_{css} &=& \left(\frac{1}{y^c} -s_2\right) \theta(1-s_2 y^c),\nn
\lim_{s_1\to 0,s_2\to 0} r_{css} &=& \frac1{y^c},\nn
\lim_{s_2\to 0} r_{css} &=& \frac{s_1}{\exp[s_1 y^c]-1},
\eea
where the first equation gives the Litim regulator for $s_2=1$ \cite{Litim:2000ci,Litim:2001up}, the second gives the power-law regulator, and the third gives the exponential one for $s_1=1$. The advantage of the css regulator is that one can continuously deform these regulator from one to another by only two parameters $s_1$ and $s_2$.

There are several possibilities to find the optimized regulator for a given model. A simple and plausible method is to find such regulator, where the calculated observable depends least on the regulator parameters. This is the so-called principle of minimal sensitivity (PMS) \cite{Canet:2002gs,Canet:2003qd,Canet:2004xe}. Unfortunately this description might suffer some systematic errors in certain models. There are nice constructive methods to get an optimized regulator \cite{Pawlowski:2015mlf}, but their usage is quite restricted, unfortunately. In LPA the Litim regulator proved to  be the optimized one, but there are only few limited results of optimization beyond LPA, and in realistic models \cite{Pawlowski:2015mlf}. This is the situation with QEG, the constructive optimization has not been performed so far even in the EH truncation. The inclusion of the coupling $b$ of the $R^2$ term in QEG makes the optimization problem much more difficult, therefore we content ourselves with a PMS investigation.

\section{Results}\label{sect:res}

We investigated the phase structure of QEG and analyzed the scaling behavior of the UV and IR regions. Then, we determined numerically the exponent $\nu$ in both regimes. The calculations differ in the two cases. In the UV regime we can perform a stability analysis around the fixed point, since the UV NGFP is a proper fixed point of the model. We should linearize the $\beta$ function around the UV NGFP, which gives a stability matrix with its eigenvalues denoted by $\theta_i$, $i=1,2,3$. We identify the critical exponent $\nu$ from the real part of the complex pair of eigenvalues $\theta_{1,2}=-\theta'\pm i\theta''$ as $\nu=1/\theta'$ ($\theta'$ is positive). The value of $\theta_3$ is negative. Due to the negative real parts of the eigenvalues the UV NGFP is an attractive fixed point. The imaginary parts of $\theta_{1,2}$ cause spiral form for the trajectories.

The stability matrix can be easily got in the vicinity of the GFP. We obtain that $\theta_1=-2$, $\theta_2=d-2$ and $\theta_3=d-4$. Generally, the GFP is a saddle point for $d>2$, i.e. it can have some attractive and some repulsive directions for the trajectories in the phase space. This is the case in QEG, too. Usually the saddle point plays the role of the phase separation by pushing the trajectories into two different regions of the phase space. We note that in scalar models the GFP is usually UV attractive making the models asymptotically free. There the Wilson-Fisher fixed point is the saddle point, which separates the trajectories belonging to the symmetric and the broken symmetric phase.

We cannot calculate the stability matrix for the IR fixed point, because the point makes the $\beta$ function and its derivatives singular. There we use another method to determine $\nu$, which is discussed in detail later.

\subsection{Ultraviolet scaling}

The QEG model with EH truncation is widely investigated. There are several works, which deal with determining the phase structure and calculating the position of the fixed points \cite{Reuter:2012id}. The property of the GFP is well known, we usually concentrate on the behavior of the UV NGFP. We are interested in the regulator dependence of the critical exponent $\nu$ for the correlation length $\xi$ in QEG.

\begin{center}
\begin{figure}[ht]
\includegraphics[width=8cm,angle=-90]{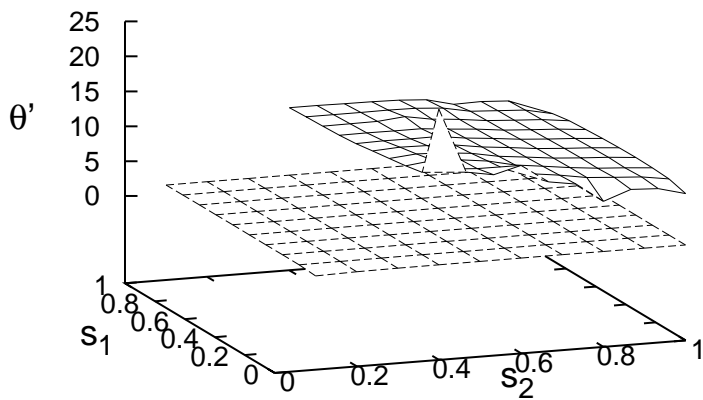}
\caption{\label{fig:theta} The regulator parameter dependence of the eigenvalue $\theta'$ is presented for the EH truncation (dashed grid) and for the $R^2$ truncation (solid grid).}
\end{figure}
\end{center}

In order to compare the results for the $R^2$ truncation, first we repeated the calculation for the exponent $\nu$ for the EH truncation in $d=3$. We used the css regulator and  calculated numerically $\theta'$ on the plane $s_1,s_2=0\dots 1$. We chose $c=1$, but we note, that in some models the power law limit for $c=1$ can contain UV divergences. The results are depicted in \fig{fig:theta}.
The calculation can be performed only for such $s_1$, which is a slightly larger than zero, since the regulator is singular at $s_1=0$. According to \fig{fig:theta} the eigenvalue shows very weak parameter dependence, all their values are around $\theta'\approx 1$. According to the precise values of $\theta'$ we obtained that the values $s_1=0$ and $s_2\approx 1$ exhibits a local minimum in the $s_1,s_2$ plane, which corresponds to the limit of the Litim regulator, where we have $\theta'=1.145$.

The evolution equations including the coupling belonging to the $R^2$ term has been derived in \cite{Lauscher:2001rz,Lauscher:2002sq,Rechenberger:2012pm}. In this work we use the equations in \cite{Lauscher:2002sq,Rechenberger:2012pm}, which can be used for arbitrary dimensions and for arbitrary regulators. Similarly to the case of the EH truncation, first we calculated $\theta'$ as the function of the regulator parameters. The results are also shown in \fig{fig:theta}. The numerical calculations were performed by a computer algebraic program. First we calculated the Litim limit of the css regulator, $s_2=1$ and $s_1=0$. We took the limit $s_1\to 0$ and obtained that at about $s_1\approx 0.001$ our numerical result provides precisely the result obtained by the Litim regulator in \cite{Rechenberger:2012pm}. This numerical result is far from trivial, since for the Litim regulator the loop integrals were performed analytically in \cite{Rechenberger:2012pm}, however for the css regulator we should calculate the loop integral numerically. As to the next we calculated $\nu$ for further values of the regulator parameters moving gradually farther away from the Litim limit.

Due to the complicated structure of the fixed point equations it is a very difficult numerical problem to solve them properly. In order to find the UV NGFP, we chose several different trial points around the GFP which were not so far from each other. Nevertheless, we should realize, that the results gave us different UV fixed points, and among them there were several fake ones. We found similar behavior in the Litim limit of the regulator, too. This fact forced us to determine the fixed point directly from the RG equations. We chose the initial value of the couplings close to the GFP and solved the RG equation numerically for increasing RG scale $k$ towards the UV region. This strategy seems to contradict the original philosophy of the RG method since there we eliminate the degrees of freedom with decreasing RG scale $k$, however mathematically this technique serves a simple trick to find the fixed point of a coupled set of differential equations.

After determining the UV NGFP and the corresponding exponent for the Litim regulator, we went on by changing the regulator parameters slightly. We assumed that small changes in the css parameters do not change the position of the fixed point too much. We stopped the evolution for that value of the RG scale $k$ which stabilizes the third digit of the position coordinate for the UV NGFP.  One can see in Fig. \ref{fig:theta}, that the exponents are calculated only at a certain part of the css parameter space. In the further regions it was not possible to locate the UV NGFP. The reason of the failure was the complicated momentum dependence of the css regulator which gave quite weak convergence of the momentum integral, and it caused extremely large computational time for that regulator parameters.

According to the philosophy of the PMS, we should find that region of the regulator parameter space where the observable shows the weakest regulator dependence. In case of the EH truncation the Litim limit gave us the optimized one. From \fig{fig:theta} the first thing that we should recognize for the $R^2$ truncation, is that the value of the exponent increases significantly. We have $\theta'\approx 10$, which is an order of magnitude larger, than the value in the EH truncation. One can see in \fig{fig:theta}, that there is a local minimum on the $s_1,s_2$ plane around the Litim limit of the regulators, $s_1\to0$ and $s_2\to 1$. The extremum can be found at $s_1=0$ and $s_2=0.8$, which corresponds a modified Litim regulator, i.e.
\beq\label{modlit}
r_{opt} = \left(\frac{1}{y} -0.8\right) \theta(1-0.8 y)
\eeq
The values of $\theta'$ in \fig{fig:theta} show larger fluctuations for the $R^2$ truncation, than in the EH one. Around $s_1=0$ and $s_2=0$ the eigenvalue seems to diverge, which might happen due to the appearing divergent integral for the power law regulator at $c=1$. Although we found the optimized regulator for different orders of the curvature $R$ and we also found the optimized value of $\theta'$, the values do not seem to converge. It might suggest that there is no convergence at all if we could add terms beyond the quadratic order in $R$.

\begin{center}
\begin{figure}[ht]
\includegraphics[width=8cm,angle=-90]{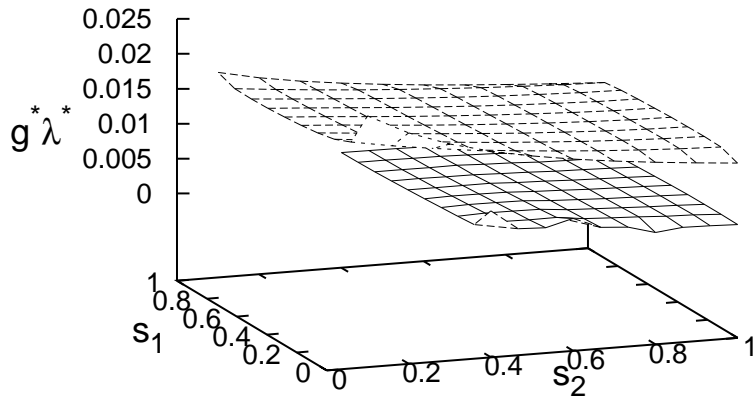}
\caption{\label{fig:gl} The product $g^*\lambda^*$ is calculated for the EH truncation (dashed grid) and for the $R^2$ truncation (solid grid).}
\end{figure}
\end{center}

As a further analysis we calculated the product of the UV NGFP coordinates $g^*\lambda^*$. Certain results showed that this quantity is less sensitive to the regulator parameters than the exponents themselves. We determined the value of $g^*\lambda^*$ for the EH and the $R^2$ truncations, too, as seen in \fig{fig:gl}.
The product shows similar regulator dependence as was obtained for $\theta'$. Instead of finding a universal quantity $g^*\lambda^*$ we can use this product for finding the optimized regulator as the value of $\theta'$. For the EH truncation we have a monotonically increasing function of $g^*\lambda^*$ for increasing $s_1$ and $s_2$. The weakest parameter dependence appears again in the Litim limit, where we obtained that $g^*\lambda^*=0.0125$. Naturally, the resulting optimized regulator agrees with the one obtained for the eigenvalue $\theta'$. The inclusion of the $R^2$ term gives smaller values for the product $g^*\lambda^*$ with about one order of magnitude. Again, it might raise the question of the convergence of the expansion in $R$. We have a local minimum at $s_1=0$ and $s_2=0.8$, which is exactly the same point in the $s_1,s_2$ plane as was found for the exponent $\nu$. Naturally we got the same a modified Litim regulator given in \eqn{modlit}. We obtained $g^*\lambda^*=0.0035$ for the optimized regulator.  For the $R^2$ truncation it seems plausible to consider some combinations of the three coordinates of the NGFP, e.g. $g^*\lambda^*b^*$ or $g^*\lambda^*/b^*$, etc., however they did not improve the regulator independence, they showed the same behavior with the same extremum found at $s_1=0$ and $s_2=0.8$, again. 

The spectral dimension for QEG can be calculated along the trajectories \cite{Lauscher:2005qz,Rechenberger:2012pm}. For the $R^2$ truncation it has poles when the coupling $\lambda$ changes its sign as the trajectory tends to the GFP from the NGFP. For the EH truncation we have positive values for $\lambda$ along the separatrix, therefore along the trajectories tending to the IR fixed point $\lambda=1/2$ the value of $\lambda$ remains positive. This is not the case for the $R^2$ truncation. For the Litim regulator it was found \cite{Rechenberger:2012pm}, that $\lambda$ can have negative values making the spectral dimension singular. The css regulator offers us the possibility to look for such regulator for which  $\lambda$ would remain positive during the evolution. It could give us a further possibility of the optimization condition for the regulators. Unfortunately we found negative $\lambda$ values for all the values of $s_1$ and $s_2$ that we could consider. However it does not rule out that other regulators can give positive $\lambda$ values, and well-defined spectral dimensions along the trajectories globally.

\subsection{Infrared scaling}

There is a point belonging to the positive IR values of $\lambda$ in the phase space which attracts the trajectories. We call this point as the IR fixed point. It is not a real fixed point, since it is not the solution of the fixed point equations, the point makes the RG equations singular. However there is a region at around that singularity point in the phase space, where a particular IR scaling regime emerges, this is the reason why we can call this point as a fixed point.

At the IR fixed point the RG flow equations loose their validity, because they become singular. There is a well defined singularity condition given by the denominators of the flow equations. The singularity condition defines the position of the IR fixed point, i.e. the point which satisfies the singularity condition. The point is situated in the positive IR values of $\lambda$ region of the phase space. In scalar models it corresponds to the broken symmetric phase. It behaves as an attractive fixed point in the IR, since it attracts all the trajectories belonging to the broken phase to a single point. To get the exponent $\nu$ of the correlation length $\xi$ we use a simple trick. We calculate the distance between the initial coupling and the separatrix. This quantity can be identified with the reduced temperature in statistical physics. This identification works perfectly in field theoretical models \cite{Nagy:2012ef}. We determine the stopping scale $k_c$, where the evolution stops if we start to solve the RG equations with the corresponding initial couplings. We choose the correlation length as the reciprocal of the stopping scale, $\xi\equiv 1/k_c$. The choice of $\xi$ is motivated by the fact that the value of $k_c$ as a certain momentum defines a characteristic distance scale in the model. Below $k_c$ the momenta are unimportant so as the larger distances in the coordinate space. It makes plausible to choose the largest distance which has physical meaning as the correlation length.

Several models show that the middling crossover fixed point drives the behavior of the IR fixed point. This is the case with QEG. There the GFP plays the role of the crossover fixed point. Fortunately it is easy to get the eigenvalues of the stability matrix, they equal to the canonical mass dimension of the couplings, from which we can find that, $[\lambda]=-2$, $[g]=d-2$, and $[b]=d-4$. The third coupling is relevant (irrelevant) for $d<4$ ($d>4$), respectively. In the case of $d=4$ the coupling $b$ is marginal, and it is expected, that it does not affect the IR behavior. For $d>4$ we have two relevant couplings, however the scaling is driven by the single irrelevant one, since its negative reciprocal gives the exponent. The most interesting case is when $d<4$, therefore we chose $d=3$. We looked for such trajectories which start from the vicinity of the GFP and tend towards the IR fixed point which is located at $\lambda^*=1/2$, $g^*=0$, and $b^*=0$ \cite{Rechenberger:2012pm}. It is an extremely difficult numerical problem, since the 3-dimensional phase space is divided into several regions with different scaling behaviors \cite{Rechenberger:2012pm}. We note that for the EH truncation we had a much easier situation. We fixed the initial value $g(\Lambda)$ and we fine tuned the value of $\lambda$ in order to approach the separatrix arbitrarily close. Here the choice of the initial coupling $g$ was arbitrary. It is for sure, that the change of $\lambda(\Lambda)$ intersects the separatrix. This numerical procedure cannot be used for the $R^2$ truncation in QEG in this simple way, since we should intersect the line of separatrix in a three dimensional phase space. For a given $g(\Lambda)$ we have a 2-dimensional plane spanned by $\lambda(\Lambda)$ and $b(\Lambda)$. In the plane the separatrix represents a single point, and the initial values $\lambda(\Lambda)$ and $b(\Lambda)$ represent another one. The latter point should be chosen in such a way that the fine tuning of the initial values approach the separatrix point arbitrarily close. The difficulty of the problem is that we do not know the exact position of the separatrix point. We note, that during the fine tuning we should give the initial couplings up to about 10 digits, therefore we should determine the position of the separatrix point in this level of accuracy. The difficulty does not come only from the fact that we have 3 couplings. This problem usually appears when we try to map out the phase structure of a saddle point, which is the case now in QEG.

\begin{center}
\begin{figure}[ht]
\includegraphics[width=8cm,angle=-90]{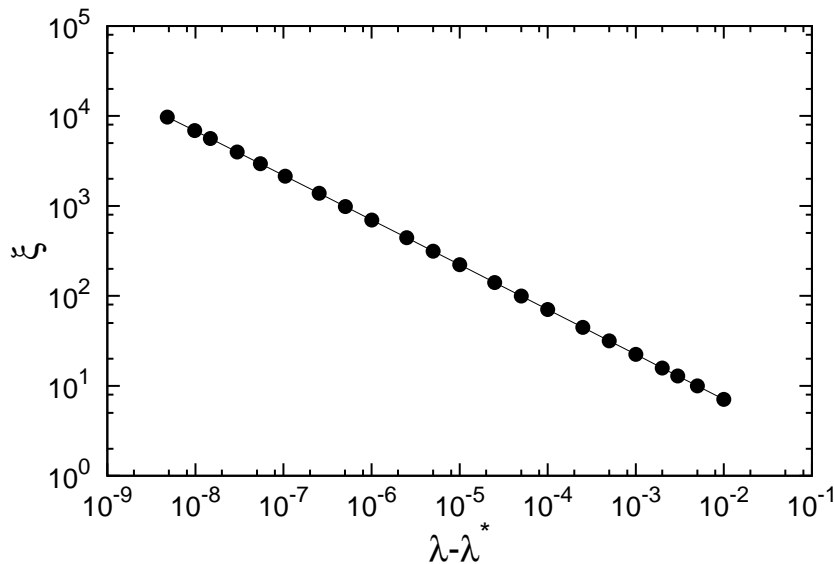}
\caption{\label{fig:ir} The correlation function $\xi$ is presented as the function of $\lambda-\lambda^*$ in log-log scale.}
\end{figure}
\end{center}

The IR region is investigated only in the case of the Litim regulator due to its high numerical complexity. First we recapitulated that there is an IR fixed point in the model at $\lambda^*=1/2$, $g^*=0$, and $b^*=0$ \cite{Rechenberger:2012pm}. It is found that if we fix the initial coupling as $b(\Lambda)=10^{-4}$ then we can approach the separatrix appropriately by tuning only the initial $\lambda(\Lambda)$. The initial values were chosen in such way that the trajectories do not intersect the singularity locus, and the fine tuning of $\lambda$ drives the trajectory to arbitrarily close to the IR fixed point. The result is presented in \fig{fig:ir}.
The exponent can be obtained as the negative slope in \fig{fig:ir}. It shows that $\nu=0.497$, which is very close to the mean field value of the exponent. We obtained the same result as for the EH truncation \cite{Nagy:2013hka}. It is quite surprising, since one might think that the irrelevance of $b$ affects the value of the exponent. It seems that this is not the case, because the irrelevance of $\lambda$ is stronger and it totally suppresses the irrelevance of $b$ during the IR scalings. For the coupling $b$ we can get stronger irrelevance, if $d<2$ which is an unimportant model. The higher order terms in the curvature can give arbitrarily strong irrelevant scalings even in $d=4$. Unfortunately the derivation of the evolution equation for those couplings are hopeless within this framework.

\section{Conclusions}\label{sect:con}

We investigated the quantum Einstein gravity extended by an $R^2$ term in the curvature in the framework of the functional renormalization group method. We mapped out the phase space of the model, and then, we numerically determined the position of the fixed points. We looked for the regulator dependence of their critical exponents. To this end we used the compactly supported smooth regulator, where there is a possibility to deform the regulator continuously among the commonly used ones. The investigation is performed in $d=3$.

There is an ultraviolet non-gaussian fixed point in the extended QEG. We determined the corresponding critical exponent of the correlation length as the function of the regulator parameters and we found, that such regulator exhibits the least sensitivity which is close to the Litim regulator.

We also investigated the IR fixed point of QEG with the $R^2$ term. Due to its complexity this calculation was done only with the Litim regulator. We found the IR fixed point, and we could determine the critical exponent of the correlation length. We found that its value is $\nu=1/2$ which is the similar result that was obtained in the absence of the coupling $b$ belonging to the $R^2$ term. The reason of the insensitivity of the exponent to the $R^2$ term can be that the coupling $b$, irrelevant in $d=3$, is less irrelevant than the coupling $g$. Naturally the inclusion of terms of the order higher than $R^2$ may overwrite our results since further couplings can appear with arbitrarily high irrelevance.

The matter of regulator dependence is an extremely important issue in the renormalization, because it guarantees the reliability of the results for the observables that are calculated. We note that these considerations for the regulator optimization can only be used in Euclidean spacetime. There are more and more attempts to formulate quantum Einstein gravity with Lorentzian signature, however the usage of the commonly used regulators could be problematic \cite{Polonyi:2017xdb}. This fact makes questionable, whether the QEG can be treated by the RG method in Lorentzian spacetime.

\section*{Acknowledgments}
S. Nagy acknowledges financial support from a J\'anos Bolyai Grant of the Hungarian Academy of Sciences, the Hungarian National Research Fund OTKA (K112233). Z. Peli acknowledges the support through the new National Excellence Program of the Ministry of Human Capacities.

\bibliography{nagy}

\end{document}